\documentstyle[supplement,epsf,wrapft]{ptptex}

\title{
The ${\rm V}_{15}$ Molecule, a Multi-spin Two-level System:\\
Adiabatic LZS Transition with or without Dissipation \\
and Kramers Theorem.
}

\author{
B. {\sc Barbara}$^{1}$, I. {\sc Chiorescu}$^{2}$, 
W. {\sc Wernsdorfer}$^{1}$, H. {\sc B\"ogge}$^{3}$, and A. {\sc M\"uller}$^{3}$
}

\inst{
$^{1}$Laboratoire de Magnetisme Louis N\'{e}el, CNRS, BP 166, 38042-Grenoble, France \\
$^2$Present address: Delft University of  Technology, Dept. of Applied Physics, 2628 CJ Delft, The
Netherlands.\\ 
$^3$Fak\"ultat f\"ur Chemie, Universit\"at Bielefeld, D-33501 Bielefeld, Germany
}

\recdate{%
}

\abst{
The so-called ${\rm V}_{15}$ molecule, of formula ${\rm K}_6 [{\rm V}_{15}^{\rm IV}{\rm As}_6{\rm 
O}_{42} ({\rm H}_2{\rm O})]8{\rm H}_2{\rm O}$, is formed of 15 spins $\frac{1}{2}$ 
antiferromagnetically coupled. The resultant spin is equal to $\frac{1}{2}$. Contrary to what is 
expected at first sight, this half-integer spin is gapped. We show that this is a consequence of the
multi-spin character of the molecule. Time reversal symmetry (Kramers theorem) is preserved due to 
the existence of a groundstate degeneracy larger than 2. Magnetization experiments have been 
performed at different temperatures, sweeping-field rates and couplings with the cryostat. Their 
interpretation in terms of the Landau-Zener-St\"{u}ckelberg  model gives evidence for adiabatic LZS 
spin reversal with and without dissipation.
}

\newcommand{\be}{\begin{equation}}
\newcommand{\ee}{\end{equation}}
\newcommand{\bea}{\begin{eqnarray}}
\newcommand{\eea}{\end{eqnarray}}
\newcommand{\vecvar}[1]{\mbox{\boldmath$#1$}}
\newcommand{\degree}{\kern-.2em\r{}\kern-.3em}
\newcommand{\degC}{\kern-.2em\r{}\kern-.3emC}

\begin{document}

\maketitle

\section{Introduction}

Contrary to high spin molecules such as ${\rm Mn}_{12}$ or ${\rm Fe}_8$ ($S=10$)\cite{ref01}, the 
${\rm V}_{15}$ molecule has a low resultant spin ($S=\frac{1}{2}$). This leads to important 
differences with respect to high spin molecules. In particular, energy barrier between spin-up and
spin-down directions, as well as dipolar couplings between molecules, become negligibly small. 
Nevertheless, in low spin molecules such as ${\rm V}_{15}$, the Hilbert space is as huge as in 
large spin molecules. This is simply because they are both multi-spins. In ${\rm V}_{15}$ the 
Hilbert space dimension is equal to $2^{15}$; in ${\rm Mn}_{12}$ and ${\rm Fe}_8$ it is $10^8$ and 
$6^8$ respectively. In principle, this molecule should be discussed in terms of the entanglement of 
the 15 spins $\frac{1}{2}$. However, considerable simplifications occur at low temperature. The 
groundstate spin is equal to $\frac{1}{2}$ and the first excited one is equal to $\frac{3}{2}$. 
These levels being reasonably well separated in energy, low temperature experiments deal with the 
groundstate only. We showed experimentally that in zero field, the symmetric and antisymmetric
superposition of the groundstates $|\frac{1}{2},S_z=\frac{1}{2}\rangle$ and 
$|\frac{1}{2},S_z=-\frac{1}{2}\rangle$ are split by an energy $\Delta$. The origin of this 
splitting with a half-integer spin is attributed to its multi-spin origin. In the absence of 
decoherence, and in zero field, the ensemble of molecules should coherently oscillate
between the symmetric and antisymmetric states $|\frac{1}{2},S_z=\frac{1}{2}\rangle\pm
|\frac{1}{2},S_z=-\frac{1}{2}\rangle$ with the frequency $\Delta/h$. In the presence of a sweeping 
field $v_H={\rm d}H/{\rm d}t$, the initial state $|\frac{1}{2},S_z=\frac{1}{2}\rangle$ either 
rotates to the state $|\frac{1}{2},S_z=-\frac{1}{2}\rangle$ or is not modified, depending on
the sweeping field rate. The switching probability depends on the comparison of the 
time $\tau_{\rm sw}=\Delta/v_Hg\mu_{\rm B}$ during which the field sweeps the mixing region 
($\sim\Delta/g\mu_{\rm B}$), with the oscillation time $\tau_{\rm os}=\hbar/\Delta$. When the field 
changes slowly ($v_{H}\ll\Delta^2/\hbar g\mu_{\rm B}$) the system stays in the same state 
(groundstate) and the spin rotates on an adiabatic way from $|\frac{1}{2},\frac{1}{2}\rangle$ to 
$|\frac{1}{2},-\frac{1}{2}\rangle$. When the field changes rapidly ($v_H\gg\Delta^2/\hbar\mu_{\rm 
B}$) the system has not the time to experience quantum mixing and the spin does not rotate. The 
switching field probability is given by $P=1-\exp(-\Gamma)$ with $\Gamma =\pi\Delta^2/2\hbar v_H 
g\mu_{\rm B}$. This is the main result of the Landau-Zener-St\"{u}ckelberg (LZS) without 
dissipation. This model was introduced for inter-band tunneling in semiconductors \cite{ref02}. It 
has been recently applied to the specific case of mesoscopic magnetism by Miyashita\cite{ref03}. 
The effects of dissipation have been considered by several authors\cite{ref04}. 

Our time resolved magnetization measurements performed at different conditions of field, 
temperature and coupling with the cryostat, are interpreted on the basis of a LZS model. Following 
Abragam and Bleaney, we introduced the role of phonons and spin bath through simple spin-phonon 
and spin-spin transitions\cite{ref05}.

After this introduction, the paper is ordered as follows: 2) Structural and thermodynamic magnetic 
properties, 3) Time resolved magnetization measurements, 4) Quantitative approach and evidence
for gapped of half-integer spin, 5) Origin of the gap in the ${\rm V}_{15}$ molecule, 6) Conclusion.

\section{Structural and thermodynamic magnetic properties}

The so-called ${\rm V}_{15}$ molecule is a complex of formula ${\rm K}_6 [{\rm V}_{15}^{\rm IV}{\rm
As}_6{\rm O}_{42} ({\rm H}_2{\rm O})]8{\rm H}_2{\rm O}$. This complex forms a lattice with  
trigonal symmetry ($a=14,02.9$~\AA, $\alpha=79.26$\degree, $V=2632$~\AA$^3$), 
containing two ${\rm V}_{15}$ complexes per cell. The third order symmetry axis of the unit-cell is 
also the symmetry axis of the ${\rm V}_{15}$ clusters (space group $R\bar 3c$). All the fifteen 
${\rm V}^{\rm IV}$ ions of spin $S=\frac{1}{2}$ are placed in a quasi-spherical layered structure 
formed of a triangle sandwiched between two non-planar hexagons (Fig.1).

There are five antiferromagnetic exchange constants. Each hexagon contains three pairs of strongly 
coupled spins (J= 800 K) and each spin at a corner of the inner triangle is coupled to two of those 
pairs (one belonging to the upper hexagon and one belonging to the lower hexagon). We have two 
different determinations of the exchange couplings shown in Fig.1: $J_1=J'=-30$~K, 
$J_2=J''=-180$~K, $J=-800$~K \cite{ref06} or $J_1=J'=-150$~K, $J_2=J''=-300$~K, $J=-800$~K 
\cite{ref07}. These values should not be taken as definitive. There are too many free parameters in 
these evaluations. However, the orders of magnitude are correct. The ${\rm V}_{15}$ molecule can be 
seen as formed by three groups of five ${\rm V}^{\rm IV}$ ions. Each group with a resultant spin 
$\frac{1}{2}$ is located on a corner of the inner triangle. These three spins$ \frac{1}{2}$ 
interact with each other through two main paths, one passing by the upper hexagon and one passing 
by the lower one. They are related to each other by the threefold symmetry axis of the molecule. 
This is a typical example of a frustrated molecule.

The magnetization measurements given below were obtained on single crystals of the ${\rm V}_{15}$ 
complex. A small dilution refrigerator allowing measurements above 0.1~K was inserted in an 
extraction magnetometer providing fields up to 16 T, with low sweeping rates of about 1 
minute/point. Below 0.9~K we clearly observed three jumps, one in zero field, and two at $H_1= \pm 
2.8$~T (Fig.2). They correspond to the $-\frac{1}{2}\Leftrightarrow\frac{1}{2}$ and 
$\pm\frac{1}{2}$ to $\pm\frac{3}{2}$ spin transitions, with respective saturation at $0.50\pm 
0.02$~$\mu_{\rm B}$ and $2.95\pm 0.02$~$\mu_{\rm B}$ per ${\rm V}_{15}$ molecule. Above 1~K, the 
observed jumps vanish because the $S=\frac{3}{2}$ level becomes thermally occupied. The 
magnetization curves are reversible and do not show anisotropy when the field is applied along 
different directions, within the accuracy of the experiments. Furthermore, experiments made on a 
single crystal or on several non-oriented single crystals were the same. So, any energy barrier 
preventing spin reversals must be quite small (less than 50~mK). The magnetization curves measured 
at equilibrium (Fig.2) are fitted using the Heisenberg Hamiltonian for three frustrated spins 
$S=\frac{1}{2}$ :
\be
H = - \sum_{\alpha=x,y,z} J_{\alpha}(S_{\alpha 1}S_{\alpha 2} + S_{\alpha 2}S_{\alpha 3} + S_{\alpha
3}S_{\alpha 1}) - g\mu_{\rm B}\vecvar{H}(\vecvar{S}_1 + \vecvar{S}_2 + \vecvar{S}_3)
\label{eqH1}
\ee
where $S_1=S_2=S_3=\frac{1}{2}$, $J_\alpha<0$ and $g=2$.

The three spins $\frac{1}{2}$ represent the resultant spin of the three groups of five spins 
(Fig.1). 
\begin{figure}[th]
 \parbox{\halftext}{
  \epsfxsize=5.5cm
  \centerline{\epsfbox{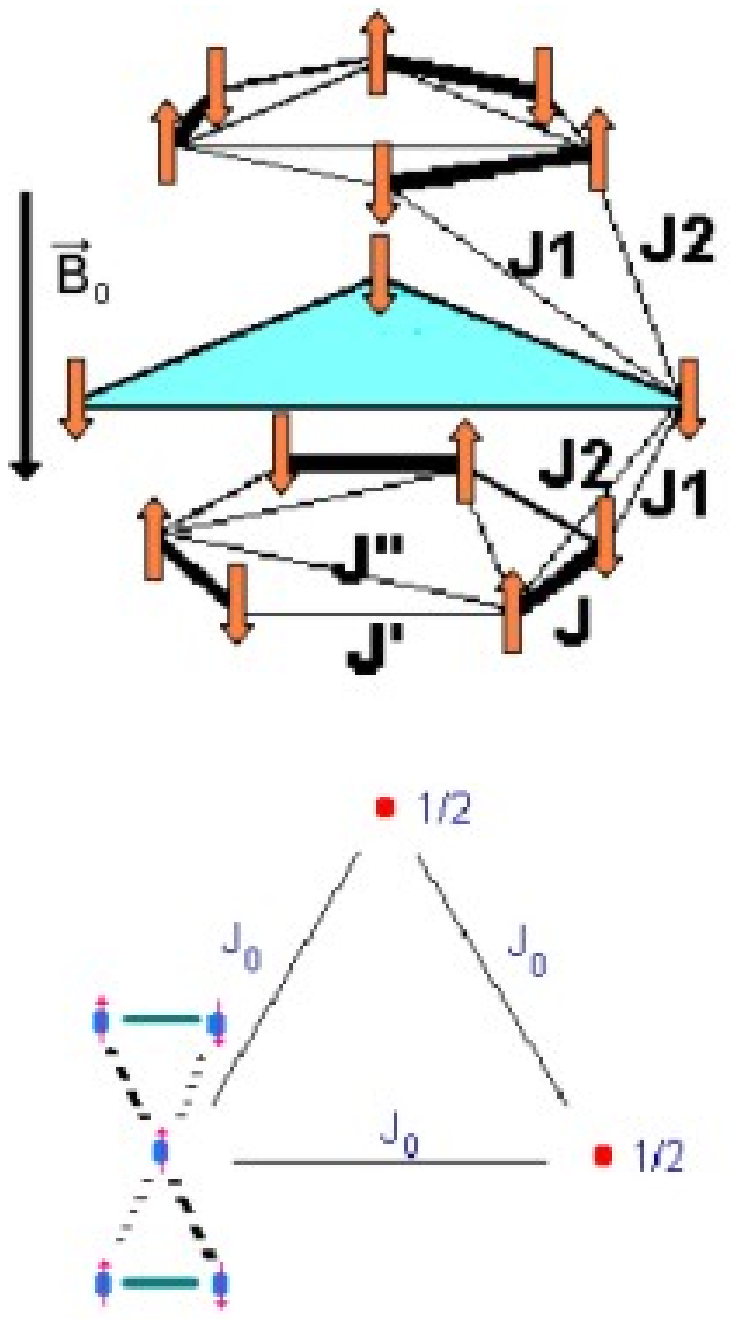}}
  \caption{Quasi-spherical layered structure formed of a triangle of ${\rm V}^{\rm IV}$ sandwiched 
between two non-planar hexagons. Main exchange paths have been indicated. Scheme of the 5 spins 
model used for our calculations.}
 \label{fig1}}
 \hspace{8mm}
 \parbox{\halftext}{
 \epsfxsize=6.6cm
 \centerline{\epsfbox{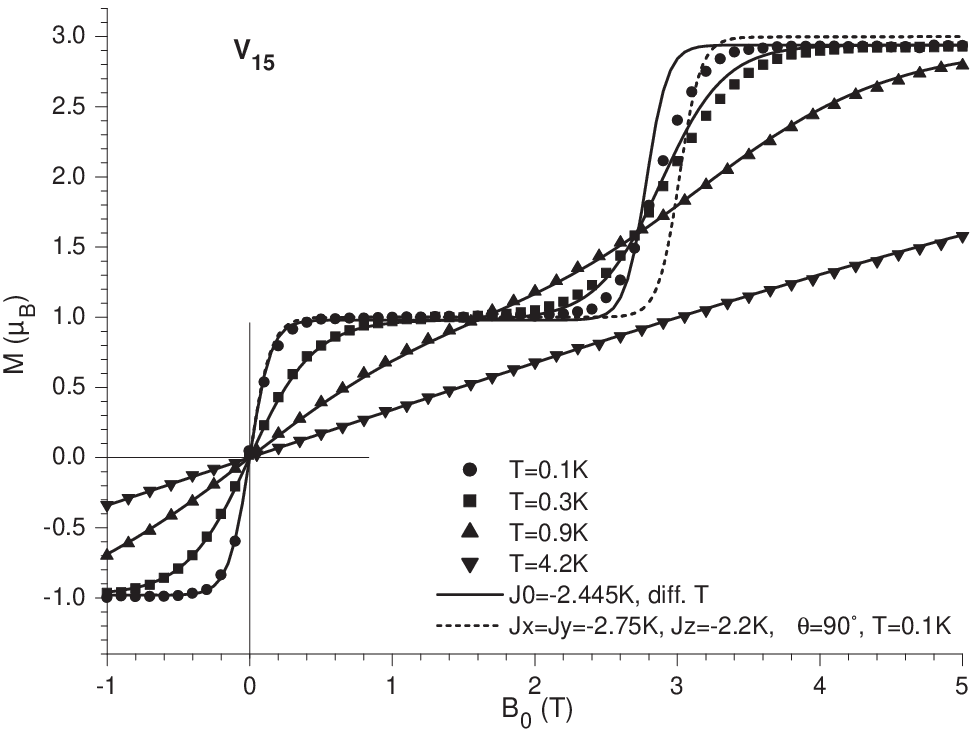}}
 \caption{Magnetization curves measured on a single crystal of ${\rm V}^{15}$. The agreement with 
the calculated curves obtained with our 5-spins model is excellent for $J_0= -2.445 {\rm K}$ 
(continuous curves). However, the spin transition measured at 2.8 T and at low temperature is
broader than the calculated one. Anisotropic exchange interactions do not influence the shape of 
the transition (dashed curve for $J_x=J_y= -2.75 {\rm K},  J_z=-2.2 {\rm K}$). This could be done 
by antisymmetric D-M interactions.}
 \label{fig2}}
 \end{figure}
The energy scheme for the lowest levels is obtained by diagonalization of the 3-spins matrix of the
Hamiltonian \ref{eqH1}. This matrix contains 64 terms, and the chosen basis is: $\{| + + +\rangle, 
| + + -\rangle, | + - +\rangle, | - + +\rangle, | + - -\rangle,  | - + -\rangle, | - - +\rangle, | 
- - -\rangle\}$. In the isotropic case and in zero field, the first excited level with spin 
$\frac{3}{2}$ has, as expected, a fourfold degeneracy. However, it turns
out that the ground state, with spin $\frac{1}{2}$, has also a fourfold degeneracy. 
This is a consequence of its multi-spin character: the groundstate of ${\rm V}_{15}$ is made 
of the juxtaposition of two independent spins $\frac{1}{2}$, each with a degeneracy of 2. We will 
see below that this ``degeneracy doubling'' effect has a general consequence on the quantum 
reversal of a spin $\frac{1}{2}$ and more generally of a half-integer spin, if interactions are 
antisymmetric.  It is easy to see analytically that the separation between the ground state and the 
first excited state is $-3J_0/2$, and that the value of the field at which the configuration 
$S=S_Z=\frac{3}{2}$ becomes favorable against the $S=S_Z=\frac{1}{2}$ is $H_1=-3J_0/(2g\mu_{\rm 
B})$.  The measured field $H_1= 2.8$~T gives $J_0\approx - 2.51$~K and then a separation
of $\approx 3.76$~K. These results are confirmed by high temperature susceptibility experiments 
(see below). The value of $\approx 3.76$~K is extremely small, compared to the exchange 
interactions $J_1, J_2, J$. This is a consequence of the spin frustration of the molecule. The 
three spins $\frac{1}{2}$ of the sandwiched triangle are coupled via the two hexagons, as indicated 
in Fig.1. According to this scheme, the exchange fields at the site of each triangle spin should 
cancel each other. The observed small value of $J_0$ results from the exchange couplings $J_1$ and 
$J_2$ not cancelling each other exactly.

This is why the separation between the lowest levels $S=\frac{1}{2}$ and $S=\frac{3}{2}$ is almost 
two orders of magnitude smaller than these interactions.  The magnetization is calculated from the 
projection of the magnetic moments upon the applied field $S_{Zi}=\langle 
i|S_{Z1}+S_{Z2}+S_{Z3}|i\rangle$, averaged over the different eigenvectors $|i\rangle$ of energies 
$E_i$. The population of different energy levels is taken at equilibrium (Boltzmann distribution). 
In the isotropic case, an analytical form of the magnetization is obtained\cite{ref08}. The fit of 
the magnetization curves measured at different temperatures gives $J_0\approx -2.445 {\rm K}$,  
which is very close to the value of 2.51 K independently obtained above. The agreement between 
calculated and measured curves is excellent at each temperature, except for the slope of the 
transition at 2.8 T which is broader in experiments. This broadening of about 0.7 T, does not 
depend on samples and cannot be ascribed to dipolar or hyperfine field distributions (about 1~mT 
and 40~mT respectively), nor to anisotropy exchange effects included in some of our calculations. 
However, we found that antisymmetric Dzyaloshinsky-Moriya interactions $\vecvar{D}_{\rm DM}$ 
($\vecvar{S}_i\times\vecvar{S}_j$) (Ref. 9 and refs. therein), which are allowed by symmetry and 
couple the states $S=\frac{1}{2}$ and $S=\frac{3}{2}$, might explain this broadening (see
section 5). In fact the important effect of Dzyaloshinsky-Moriya interactions for the 
induction of tunnel splittings, otherwise forbidden, was pointed out in \cite{ref10}
for the case of the ${\rm Mn}_{12}$ molecule. Detailled calculations have been done in  
\cite{ref11}. 

The susceptibility is calculated by taking the first derivative of the analytical expression of the 
magnetization\cite{ref08}. The zero-field susceptibility $\chi_i$ has a very simple expression:
\be
\chi_i  = \frac{(g\mu_{\rm B})^2}{4k_BT}\frac{5+e^{-3J_0/2kT}}{1+ e^{-3J_0/2kT}}
\label{eqxi}
\ee
The limits $T \ll 3J_0/2k_B$ and $T\gg 3J_0/2k_B$ correspond respectively to a spin $\frac{1}{2}$ 
with the effective moment  $\sqrt{3}g\mu_{\rm B}/2$, and 3 spins $\frac{1}{2}$ with the effective 
moment  $3g\mu_{\rm B}/2$. The fit of the susceptibility curves measured between 0.1~K and 100~K is
given Fig.3 and Fig.4. The measured reciprocal susceptibility  \emph{vs.} temperature and its 
comparison with Eq.(2) for  $J_0\approx -2.445$~K (determined from $M(H)$) is excellent. The 
straight line represents the Curie-Weiss law with the Curie constant $C=0.686$~$\mu_{\rm B}$K/T and
the paramagnetic temperature $\theta = - 12$~mK. This Curie constant corresponds to the effective 
paramagnetic moment $1.75$~$\mu_{\rm B}$, a value very close to $g\mu_{\rm B} (S(S+1))\frac{1}{2} = 
g\mu_{\rm B}\sqrt{3}/2$ for $S=\frac{1}{2}$. The paramagnetic Curie temperature is of the order of 
dipolar couplings between $S=\frac{1}{2}$ molecule spins. Above 0.5~K, the susceptibility deviates 
from the experimental data and from the calculation from Eq.\ref{eqxi}. This is because the spin 
$\frac{1}{2}$ ground-state starts to be deteriorated. Fig.4 shows the measured effective moment 
$\mu_{\rm eff}= \sqrt{3kT\chi_i}$ as a function of temperature. It increases from the value of 
$1.75$~$\mu_{\rm B}$ corresponding to the spin $\frac{1}{2}$ ground state, and tends asymptotically 
to the value of $3$~$\mu_{\rm B}$ which is preserved up to 100~K (three independent spins 
$\frac{1}{2}$). The fit of these susceptibility measured between 0.1 and 125~K to Eq.\ref{eqxi} is 
excellent with $J_0 = -2.445$~K.
\begin{figure}[th]
\parbox{\halftext}{ \epsfxsize=6.6cm \centerline{\epsfbox{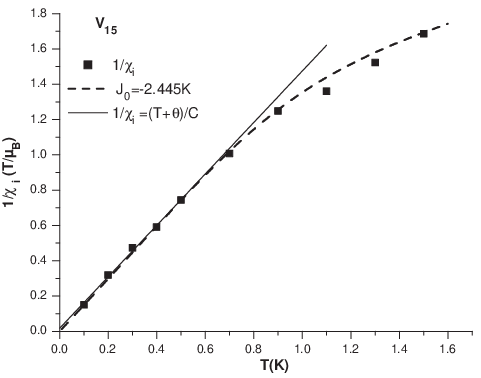}}
\caption{Measured and calculated reciprocal susceptibility \emph{vs.} temperature. The dashed curve 
comes from the model with $J_0= -2.445$~K and the continuous line is a Curie-Weiss fit 
with a paramagnetic Curie temperature $\theta = 12$~mK and a Curie constant corresponding to an 
effective moment of $1.75$~$\mu_{\rm B}$  ($S=\frac{1}{2}$).}
\label{fig3}}
\hspace{8mm}
\parbox{\halftext}{\epsfxsize=6.6cm \centerline{\epsfbox{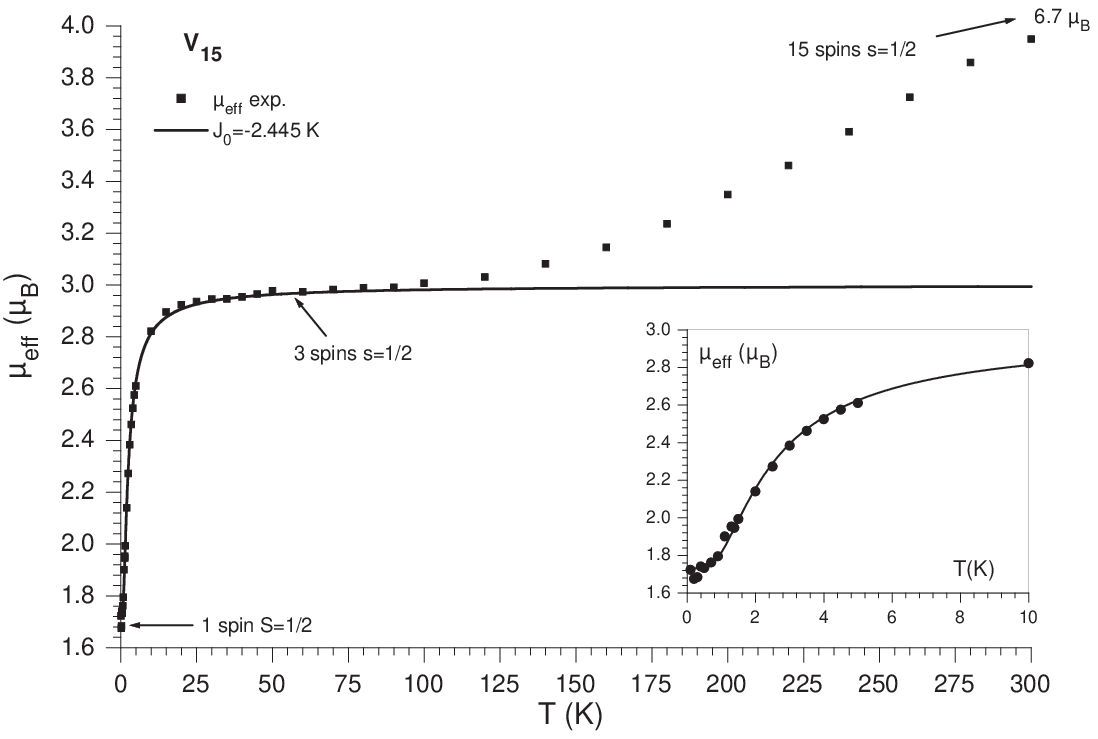}}
\caption{Curve calculated in the 5-spins model is valid up to 125~K, showing that
excited levels corresponding to triangle-hexagons couplings are well separated from the
triangle levels. Two plateaus can be seen:  one at $1.75$~$\mu_{\rm B}$ corresponding to one spin 
$\frac{1}{2}$ (below 0.5~K) and one at $3$~$\mu_{\rm B}$  corresponding to three spins 
$\frac{1}{2}$ (above 25~K).}
\label{fig4}}
\end{figure}

\section{Study of the two-level system realized in the ${\rm V}_{15}$ molecule}
 
In this section we concentrate on the zero-field $|\frac{1}{2},S_z=\frac{1}{2}\rangle\Leftrightarrow
|\frac{1}{2},S_z=-\frac{1}{2}\rangle$ transition. We studied the temperature and time dependence of 
a small single-crystalline grain of ${\rm V}_{15}$ ($L\approx 30-40$~$\mu$m), with different 
couplings to the cryostat. A first comparison with the LZS model shows that with
zero-field splitting of the order of  50-200~mK (see section 4) the transition is purely 
adiabatic if  $v_H\ll10^{10}$~T/s. In our experiments where $v_H\le 0.14$~T/s (see below) it is 
\emph{a fortiori} the case. Our experiments being done at finite temperatures, we found a much
richer behavior than simple adiabatic LZS, with two limiting cases in which the magnetization 
curves $M(H)$ are reversible, and an intermediate one where it is irreversible. This behavior 
results from effects of the environment on the adiabatic LZS model.

\subsection{Fast sweeping rate / weak coupling to the cryostat, and adiabatic LZS.}

The first limit of fast field variations and weak coupling to the cryostat, was reached by
inserting a plastic sheet, much thicker than the sample, between the sample and the sample-holder 
and restricting the temperature to lowest values. The four $M(H)$ curves of Fig.5 show 
either a weak hysteresis or no hysteresis at all, depending on the competition of temperature 
(giving hysteresis through spin-phonons transitions) and sweeping field rate $v_H={\rm d}H/{\rm 
d}t$ (preventing these transitions). This effect can be accounted for by the ratio $\alpha v_H/T^2$ 
(see the model below).  
For $\alpha v_H/T^2 < 14$~T/sK$^2$ weak hysteresis is observed on the $M(H)$ curve. It is 
attributed to finite but  small probability for spin-phonon transitions. For $\alpha v_H/T^2 > 
14-40$~T/sK$^2$ the $M(H)$ curve becomes reversible. Spin-phonons transitions have
not the time to take place and the probability for their occurrence is negligibly small; only the 
groundstate is occupied. The $|\frac{1}{2},S_z=\frac{1}{2}\rangle$ to 
$|\frac{1}{2},S_z=-\frac{1}{2}\rangle$ LZS transition is adiabatic. The adiabatic $M(H)$ curve is 
nicely fitted with the expression $M=(\frac{1}{2})(g\mu_{\rm B})^2H/[(\Delta_0)^2+(g\mu_{\rm
B}H)^2]^{\frac{1}{2}}$ obtained from $E=\frac{1}{2}{\rm d}\Delta_H/{\rm d}H$ where 
$\Delta_H=[(g\mu_{\rm B}H)^2 + (\Delta_0^2)]^{\frac{1}{2}}$ 
is the two-level field-dependent splitting energy. The saturation $(\frac{1}{2})g\mu_{\rm B}$ is 
very close to $1 \mu_{\rm B}$ as expected, and the initial slope $M/H =(g\mu_{\rm B})^2/2\Delta_0$ 
is proportional to $1/\Delta_0$, giving immediately an order of magnitude for the zero-field 
splitting $\Delta_0\approx 0.1$~K.  The fit of the whole $M(H)$ curve, as well as the $1/H^2$ 
approach to saturation gives $\Delta_0\approx 80$~mK. 

In short, a single crystal of ${\rm V}_{15}$ molecules with reduced couplings to the cryostat, is a
realization of the adiabatic LZS model applied to magnetism \cite{ref03}. The spin temperature can 
then reach very low values $T_s=T_i/(1+(g\mu_{\rm B}H/\Delta_0)^2)\approx T_i/50$, where $T_i$ is 
the initial temperature. The same procedure applied to the $\frac{1}{2}$ to $\frac{3}{2}$ spin 
transition which occurs in a rather high field (2.8~T) should allow to obtain high nuclear spin 
polarisation.

\begin{figure}[ht]
 \parbox{\halftext}{ \epsfxsize=6.6cm \centerline{\epsfbox{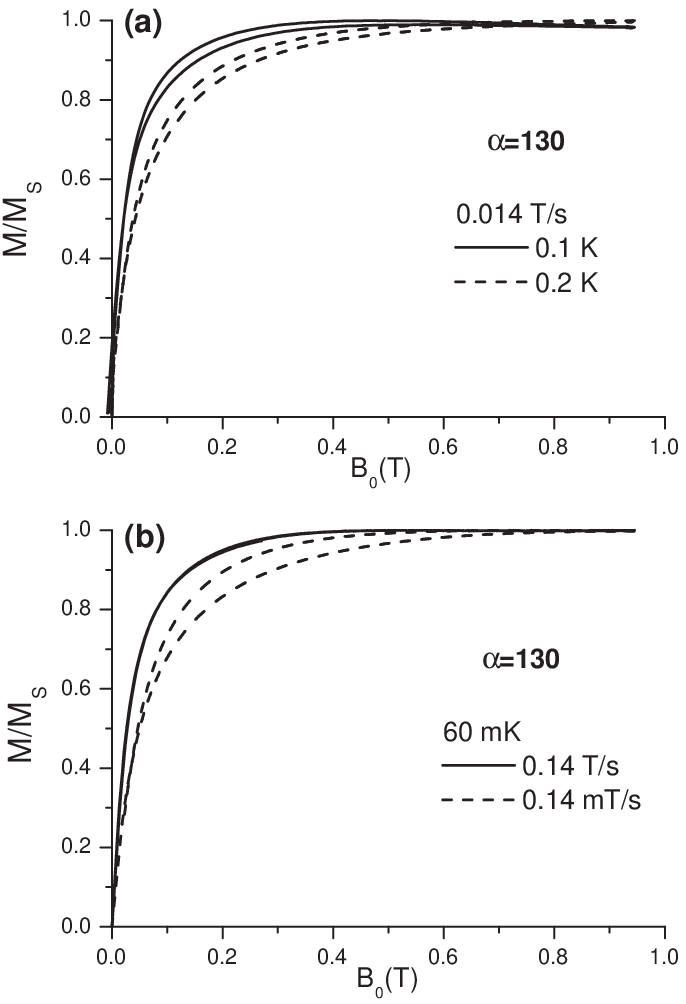}} 
 \caption{Hysteresis loops measured in the case of a sample rather well
isolated from the cryostat ($\alpha =130$~sK$^2$). The curve at 60~mK and 0.14~T/s shows an
adiabatic LZS spin transition. It is very well fitted with the expression
$\frac{M}{M_s}=\frac{{\rm d}\Delta_H}{{\rm d}H}$.}
 \label{fig5}}
 \hspace{8mm}
 \parbox{\halftext}{\epsfxsize=6.6cm \centerline{\epsfbox{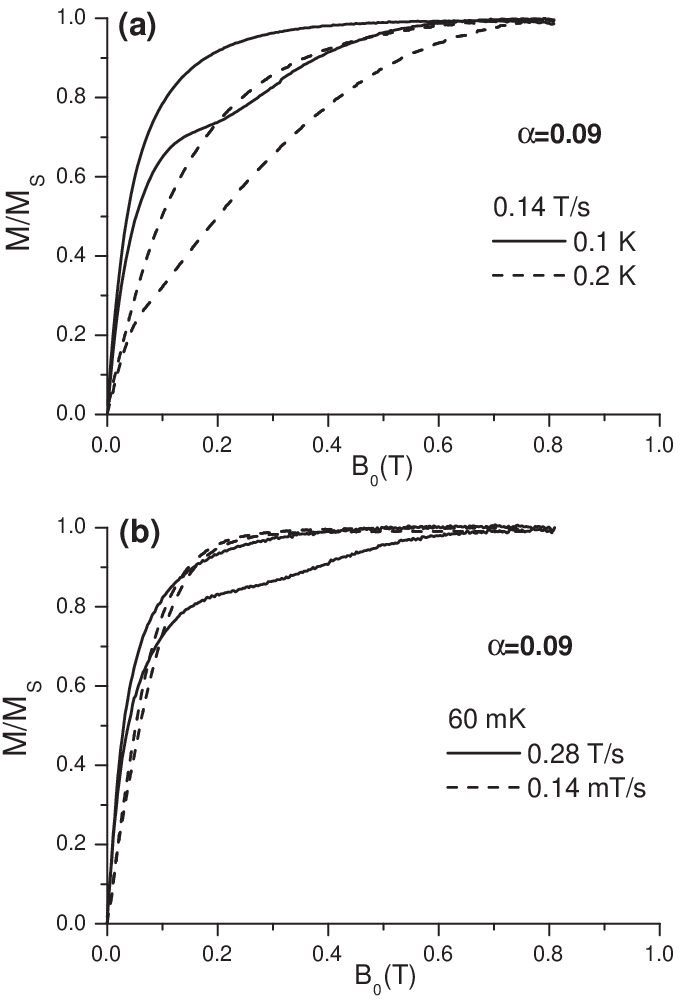}}
 \caption{Hysteresis loops measured in the case of a sample rather well
coupled to the cryostat ($\alpha = 0.09$~sK$^2$).}
 \label{fig6}}
 \end{figure}

\subsection{Low sweeping rate / strong coupling to the cryostat, and dissipative LZS.}

Here the coupling between the sample and the cryostat is made better in order to favor spin-phonon
transitions. We considered two cases. In the first one, the coupling was obtained by simple contact 
using some grease between sample and sample holder. In the second case the coupling was better, the 
contact being made of a mixture of silver powder with grease. In both cases non-equilibrium 
hysteresis loops are observed at fast-sweeping rate, down to the lowest measured temperature of 
50~mK. The $M(H)$ curves obtained with the first type of coupling (grease contact), can be found in 
Ref. 8, 9, 13. Those obtained with the best coupling with the cryostat (Ag+grease) are similar. 
They are shown Fig.6. When the field increases, coming from negative values, the magnetization 
passes through the origin of the coordinates, reaches a plateau and then approaches saturation. 
This leads to a winged hysteresis loop characterized by a reversible but temperature
and sweeping field-dependent zero-field susceptibility. The irreversible wings 
depend also on temperature and sweeping field. The sweeping field dependence of the zero-field 
susceptibility is typical of the phenomenon studied here. 

In Fig.6 the hysteresis of the $M(H)$ curves disappears when the field is slow enough. This is a 
natural consequence of thermal equilibrium. Spin-phonon transitions are sufficiently probable so 
that the two-level populations equilibrate with the Boltzmann distribution. As an example, 
equilibrium is almost reached at 60~mK and $v_H < 0.14$~mT/s (approximated by the median of the two 
branches of the $M(H)$ hysteresis loop, Fig.6). This reversibility is very different from the one 
discussed in 3.1 where field variations were fast enough to suppress spin-phonons transitions 
completely, giving a highly non-equilibrium groundstate at finite temperature (adiabatic LZS). 

In order to give a first qualitative analysis of these ``butterfly hysteresis loops''\cite{ref09} 
we will see how the level occupation numbers vary when sweeping the external field (Fig.7). The 
spin temperature $Ts$  is such that $n_1/n_2 = \exp(\Delta_H/k_{\rm B}T_S )$ where $n_{1,2}$ 
($n_{1,2{\rm equ}}$) is the out-of-equilibrium (equilibrium) level occupation numbers. The 
magnetization curves at 60~mK (Fig.~6), show a spin temperature significantly lower than the bath 
temperature ($n_1 > n_{1{\rm equ}}$, $T_S < T$) between 0.3~T and 0.15~T (the field
at which the magnetization curve intersects the equilibrium one). After this intercept, $T_S$ is 
larger than the bath temperature ($n_1 < n_{1{\rm equ}}$, $T_S > T$) and at sufficiently high 
fields (about 0.5~T) it reaches its equilibrium value ($n_1 = n_{1{\rm equ}}$, $T_S = T$).  
Note that the magnetization curves measured between 0.7 and 0.02~T at fast sweeping rates 
are nearly the same, suggesting weak exchange with the bath, \emph{i.e.} nearly adiabatic 
demagnetization. This is because below 0.5 K, the heat capacity of phonons $C_{\rm ph}$ is very 
much smaller than that of the spins $C_S$, so that the phonons temperature $T_{\rm ph}$ very 
rapidly adjusts to that of the spins $T_S$. The spins and the phonons can be seen as a single 
system. The coupling of this system with the cryostat is weak because energy transfers through 
phonons must obey the drastic condition $\hbar\omega=\Delta_H$ and the number of phonons 
available at this energy $\hbar\omega$ is extremely small: 
$n_T = \int\sigma(\omega){\rm d}\omega/\exp(\hbar\omega/kT)$. The integral is taken over the level 
linewidth $\Delta\omega$ (due to fast hyperfine fluctuations), $\sigma(\omega){\rm d}\omega= 
3V\omega^2{\rm d}\omega/2\pi v_{\rm ph}^3$ is the number of phonon modes between $\omega$
and $\omega+{\rm d}\omega$ per molecule of volume $V$, and $v_{\rm ph}$ is the phonon velocity. 
Taking the typical values for ${\rm V}_{15}$-like systems $v=3000 {\rm m/s}$, and 
$\Delta\omega=10^2$~MHz, we find at T=0.1 K, $n_T\approx 10^{\sim 6}-10^{\sim 8}$ 
phonons/molecule. The number of such lattice modes being much smaller than the number of spins, 
energy transfer between phonons from the cryostat and spins must be very difficult, a phenomenon 
known as the phonon bottleneck\cite{ref05,ref12}. Despite a good thermal contact between the sample 
and the cryostat, the energy flow from the latter is not sufficient to compensate the lack of 
phonons at energies $\hbar(\omega\pm\Delta\omega)$. Most molecules are out of equilibrium, at a 
temperature $T_S$ different from the cryostat temperature $T$.
\begin{figure}
 \epsfxsize=6.6cm 
 \centerline{\epsfbox{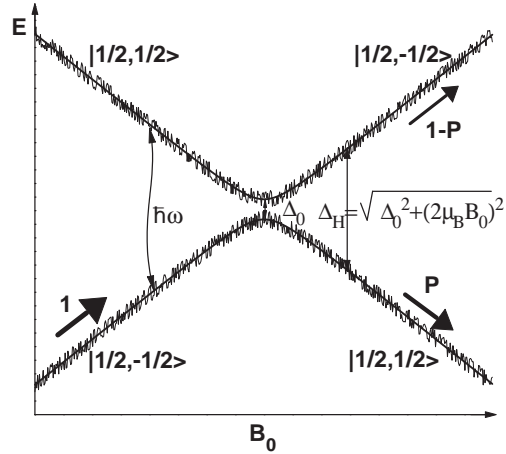}}
 \caption{Schematic representation of the ${\rm V}^{15}$ spin $\frac{1}{2}$ energy scheme in a
static magnetic field (the level broadening is due to nuclear spin fluctuations). Probability 
transitions $P$ in the LZS model and spin-phonon transitions are indicated by different types of 
arrows. The existence of a zero-field splitting $\Delta_0$ with a spin $\frac{1}{2}$ is related to 
the fact that in ${\rm V}^{15}$, this scheme is doubly degenerated with the superposition of two 
independent and identical schemes.}
 \label{fig7}
 \end{figure}

\section{Quantitative approach and numerical calculations}

In a spin $\frac{1}{2}$ two-level system, the populations of spins and phonons obey a set of two 
differential balance equations (\emph{e.g.}, see Ref. \cite{ref05}). As discussed above, at low 
enough temperature, the condition $b=C_s/C_{\rm ph}\gg 1$ implies $T_{\rm ph}\rightarrow T_s\ne T$. 
The phonon population is dragged by that of the spins, and the problem reduces to a single 
equation. 
Writing $x=(n_1-n_2)/( n_{1{\rm eq}}-n_{\rm eq})$, the spin population is given by $\tau_{H}{ \rm 
d}x/{\rm d}t =1/x - 1$. The solution of this simple equation is $-t/\tau_{H}=x-x_0 + 
\ln[(x-1)/(x_0-1)]$ where $\tau_{H}=b\tau_{\rm ph}=(\alpha/\Delta_H^2)\tanh^2(\Delta_H/2k_{\rm 
B}T)$ is the relaxation time of the spin-phonon system to the cryostat temperature and $x_0$ the
initial spin population. The constant $\alpha$ is formally proportional to the size $L$ of the 
sample, \emph{i.e.} to the distance the phonons have to travel to reach the cryostat temperature 
\cite{ref09}. In experiments, it is common to change $L$ artificially by modifying the thermal 
coupling at the sample/sample-holder interface (see above). The above expressions of
$x$ and $\tau_H$ allow to calculate the magnetic relaxation and the magnetization curves $M(H,T,t)$ 
with the use of only two parameters $\Delta_0$ and $\alpha$. Note that $\tau_H{\rm d}x/{\rm d}t$ 
could be written $\tau_Hv_H({\rm d}x/H)$, suggesting that the parameter $\tau_Hv_H\propto\alpha 
v_H/T^2$ is relevant for $M(H)$ curves.

The $M(H)$ curves have been calculated to give the best agreement with the measured ones. The 
obtained parameters are $\Delta_0\approx 50$~mK and $\alpha\approx0.15$~sK$^2$ (grease contact). 
The comparison between the magnetic relaxation curves measured in the same conditions
and calculated (not shown) gives $\Delta_0\approx 150$~mK. In the case of the best coupling between 
sample and cryostat , the agreement between measured and calculated $M(H)$ curves gives 
$\Delta_0\approx 80$~mK and $\alpha\approx 0.09$~sK$^2$. This smaller value of $\alpha$ is 
consistent with a better contact to the cryostat. These determinations of $\Delta_0$ must be 
compared to the value $\Delta_0\approx 80$~mK obtained from the fit of $M(H)$ without dissipation 
(section 3.1). Relaxation time measurements being less accurate than $M(H)$ fits, we can consider 
that the zero-field splitting of ${\rm V}_{15}$ is $\Delta_0\approx 80\pm 20$~mK.

\begin{wraptable}{l}{\halftext}
  \caption{Relaxation times at different fields and temperatures, obtained from the fit of the 
curves shown Fig.8 ($\tau_{H {\rm exp}}$) and calculated from the time dependent equation
given in the text with $\alpha=130 {\rm sK^2}$ and $\Delta_0=80 {\rm mK}$.}
  \label{table}
  \begin{center}
    \begin{tabular}{llll} \hline\hline 
   $T$(K) & $H$(mT) & $\tau_{H {\rm exp}}$(s) & $\tau_{H {\rm calc}}$(s) \\ \hline
   0.05  & 14   & 1507 & 8716   \\
   0.15  & 14   & 551 & 1323   \\
   0.05  & 70   & 3883 & 3675   \\
   0.15  & 70   & 970  & 997   \\ \hline
   \end{tabular}
  \end{center}
\end{wraptable}
Although these quantitative determinations of $\Delta_0$ are not direct (neutron scattering 
experiments are in progress, see also Ref.14), the obtained finite value is qualitatively proved by 
the sweeping 
field dependence of the initial susceptibility and in particular by the clear observation of an 
adiabatic LZS regime, in fast fields. Quantitatively, the value of about 80 mK seems rather safe 
for the following reasons. We obtained three couples ($\Delta_0, \alpha$) independently from 
the comparison of measured and calculated $M(H)$ curves in the adiabatic LZS regime without 
dissipation (bad coupling with the cryostat, $\alpha\approx 130$~sK$^2$) and with dissipation (good 
and very good couplings, $\alpha\approx0.15$ and $0.09$~sK$^2$). The $\Delta_0$ values obtained are 
close to each other and the values are consistent with the whole physical picture, giving credit to 
our approach. Furthermore, our model is self-consistent. In the case of $\alpha\approx 0.15$~sK$^2$ 
we have found a relaxation time of about 1~s, from the above expression of $\tau_H$, which is close 
to the measured one\cite{ref13}. Moreover, we obtained (i) a sound velocity $v_{\rm ph}\approx 3000 
{\rm m/s}$, which corresponds to the values given for insulating systems\cite{ref05}, and (ii) the 
calculated specific heat ratio $b =\tau_Hv_{\rm ph}/L\approx 10^8$ is comparableto the expected 
value for an ensemble of spins $\frac{1}{2}$ in a low temperature bottleneck 
regime\cite{ref05,ref12}. Finally, in the other limit  where $\alpha\approx 130$~sK$^2$ and 
$\Delta_0\approx 80$~mK, our magnetic relaxation measurements are well fitted to the model given 
above and the obtained relaxation times are very close to the calculated ones for the whole range 
of fields, excepted near zero field  where the measured relaxation (Fig.8) is about six times 
faster (Table \ref{table}). This shows that when the relaxation takes place within the zero-field 
splitting $\Delta_0$, the phonons bath is no longer sufficient to explain the dissipation. An 
excess relaxation probably comes from the Dzyaloshinsky-Moriya interactions and the spin bath, in 
particular from the fast fluctuations of the $^{51}{\rm V}$ nuclear spins.  A more detailed study 
of the interplay between nuclear and electronic spins, in the presence of the spin mixing of the 
Dzyaloshinsky-Moriya interactions should be done in the case of this molecule, in the light of our 
recent works on ${\rm H}_0^{3+}$ ions\cite{ref14}. For a discussion see the next section.

\begin{figure}
 \epsfxsize=13.2cm
  \centerline{\epsfbox{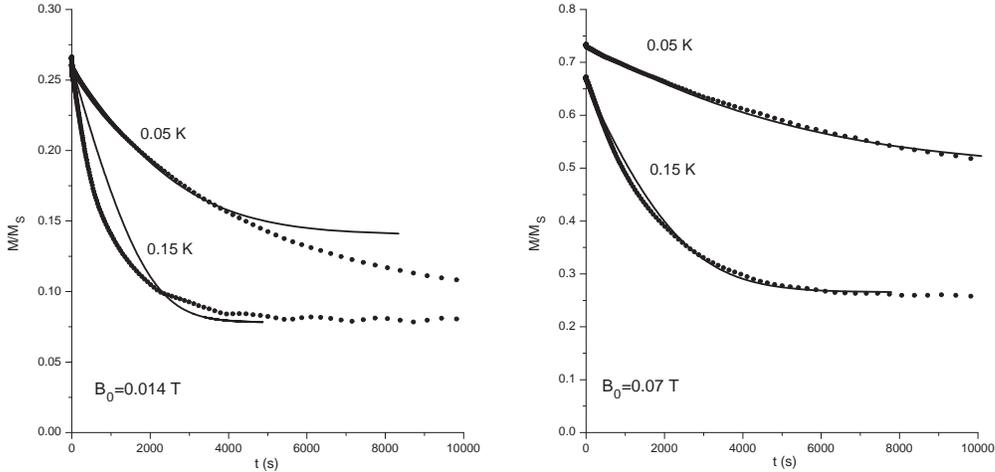}}
  \caption{Relaxation of an isolated sample, in a field $H\ll\Delta_0$ for which spin up and spin
down states are mixed (left), and in a field $H\gg\Delta_0$ where the level structure is dominated 
by the Zeeman effect.  The fits (continuous curves) of data (dots) are good except when the field 
and the temperature are small (0.15~K, 0.014~T). }
\label{fig8}
\end{figure}

\section{Origin of the gap in the half-integer $\frac{1}{2}$ spin of the ${\rm V}_{15}$ molecule}
 
It is well known that time reversal symmetry prevents half-integer spins to be gapped in zero field 
(Kramers theorem\cite{ref15}). This is also the reason for the absence of Haldane gap in magnetic 
chains with half-integer spins\cite{ref16}. Our first evidence for a gap in spins $\frac{1}{2}$ 
${\rm V}_{15}$ molecules is at first sight in contradiction with this first principle. The 
difference between the spin $\frac{1}{2}$ of ${\rm V}_{15}$ and an atomic spin $\frac{1}{2}$ is 
obviously the multi-spin character of ${\rm V}_{15}$. The groundstate spin $\frac{1}{2}$ is the 
resultant of exchange interactions as shown section 2. As we suggested in Ref. 9 and references 
therein, antisymmetric Dzyaloshinsky-Moriya (D-M) interactions can provide a finite gap in a 
multi-spin system with resultant half-integer spin. On the point view of symmetry, the molecule 
${\rm V}_{15}$ is a good candidate to show this effect, due to the lack of inversion center. This 
is particularly true for spin pairs situated near the borders of the molecules. The Hamiltonian 
\ref{eqH1}  should therefore contain the additional term:
\be         
H_{DM} = -\sum_{ij} \vecvar{D}_{ij} \vecvar{S}_i\times\vecvar{S}_j,
\label{eqHdm}
\ee
the summation being done over all spin pairs. The norm of the vector $\vecvar{D}$ can be evaluated 
from the measured anisotropy of the g-factor. Very recent EPR experiments by Ajiro et 
al.\cite{ref17} made accurate measurements of the angular variations of this factor, in a plane 
containing the c-axis. The results, much more complete  than the previous ones\cite{ref06,ref07}, 
give at 2.4~K, $g_{||c} =1.98$ and $g_{\perp c}=1.968$. Then using $D\approx J\Delta g/g$, with
$\Delta g/g =(g_{||c} -g_{\perp c})/g =0.6\%$,  we can see that the value $DS^2\approx D/4\approx 
80 {\rm mK}$ derived from measured $\Delta_0$, gives $J\approx 60 {\rm K}$, which is of the right 
order of magnitude for average exchange interactions. Therefore D-M interactions could really be
at the origin of the observed zero-field splitting.

The best theoretical proof for that was given recently by Miyashita and Nagaosa\cite{ref18} who 
calculated the magnetization curves and splittings resulting from level mixings in the 3-spins 
Heisenberg model  (Hamiltonian \ref{eqH1} in section 2) with various types of perturbations of the 
form 
\be
 H_1 = -\sum_{i,j} {\alpha}_{ij} S_{iz}S_{jx}.
\label{eqHpert}
\ee

They showed that the groundstate is not gapped, as expected from the Kramers theorem\cite{ref15}, 
only if the $\alpha_{ij}$ show the reflection symmetry, \emph{i.e.} $\alpha_{ij}=\alpha_{ji}$. In 
other cases, when $\alpha_{ij}\ne\alpha_{ji}$ the perturbation in Eq.\ref{eqHpert} creates a gap 
between the $\pm1/2$ spin states, as observed experimentally. However, as discussed above, the spin 
$\frac{1}{2}$ being a multi-spin system, its groundstate degeneracy is larger than 2. In the 
present case of 3 spins $\frac{1}{2}$ it is equal to 4. The four states with $S=\frac{1}{2}$ form 
two sets of avoided level crossing at zero field, each with two-fold degeneracy. This allows 
Kramers time reversal symmetry to be preserved even in the presence of a zero-field splitting. 
Coming back to Eq.\ref{eqHdm}, one can write $H_y=-D_{ijy} (S_{ix}S_{jz}- S_{iz}S_{jx})$, which 
identifies to $H_1$ in Eq.\ref{eqHpert} if $\alpha_{ij} = - \alpha_{ji} = D_{ijy}$, which is not 
surprising as D-M interactions are antisymmetric. 

Beyond the case of the ${\rm V}_{15}$ molecule, D-M interactions, which are very general with 
molecules (and in particular if they are of finite size), should lead to a gapped groundstate in 
non-integer multi-spin molecules. It is possible that other mechanisms, such as the 
multi-exchange couplings in spin liquids, could lead to similar effects. The molecule ${\rm 
V}_{15}$ can be considered as an example of nanometer scale spin-liquid.

Before ending this section, let us consider some characteristics \cite{ref19} of the nuclear spins 
of $^{51}{\rm V}$. The natural abundance of this element with nuclear spin $\frac{7}{2}$, is 
99.76\%. The resonance frequency and magnetogyric ratio of 2.63 GHz  and 7.045 rad/$10^7$~Ts give 
an hyperfine field of about 40~T. The inverse hyperfine field acting on electronic spins, roughly 
2000 times smaller, is in the 20~mK scale, \emph{i.e.} of the order of $\Delta_0\approx 80$~mK. As 
indicated before, this field will give a broadening of electronic levels. This broadening 
corresponds to the incoherent bunching of 8 electro-nuclear levels. Another origin for level 
broadening comes from the spin mixing by Dzyaloshinsky-Moriya interactions between the different 
pairs of spins in the molecule. The observation of faster relaxation near zero field, suggests that 
fluctuations on the two levels intercept each other near zero field (at the energy scale of 
$\Delta_0$). This is the case for nuclear spin fluctuations, the broadening of which is of the 
order $\Delta_0$/2. These fluctuations should therefore be relevant to explain the excess of 
relaxation in zero field \cite{ref08,ref09}. Level broadening, resulting from incoherent 
nuclear spins or Dzyaloshinsky-Moriya fluctuations, can be considered as a magnetic external noise 
on the LZS model\cite{ref20}. Such a noise cannot explain the origin of $\Delta_0$, in particular 
because it cannot, by itself, simulate the LZS model, nor the associated dynamics in the presence
of spin-phonon transitions. As shown above the origin of $\Delta_0$ is clearly associated with the 
static antisymmetric effect of Dzyaloshinsky-Moriya interactions.

\section{Conclusion.}

The ${\rm V}_{15}$ molecule, a multi-spin system with spin $\frac{1}{2}$, shows clearly adiabatic 
LZS transitions with or without dissipation, depending on the sweeping field rate and on the 
coupling with the cryostat. This allows to study environmental effects on quantum spin reversal in 
a two level system. In particular it is shown that the observed magnetic relaxation is associated 
with spin-phonon transitions in the Zeeman regime (phonon bath). Near zero field, level broadening 
due to spin fluctuations (essentially nuclear spin fluctuations) seem to overlap, giving an excess 
of relaxation, associated with the spin bath. An important consequence of this study is that, 
contrary to what is expected at first sight, this half-integer spin is gapped in zero-field. This 
is a consequence of the multi-spin character of the molecule plus Dzyaloshinsky-Moriya 
interactions. Time reversal symmetry is preserved due to a fourfold degeneracy of the groundstate.

Acknowledgements: B.B. would like to thank the organizing committee of the 16$^{th}$ 
Nishinomiya-Yukawa Memorial Symposium for their kind invitation, and particularly Professors T. 
Tonegawa, H. Takayama, and S. Miyashita. We would like to thank them and especially Professors S. 
Miyashita and P.C.E. Stamp for discussions on our common field of interest.


\begin{thebibliography}{99}

\bibitem{ref01} I.~Tupitsyn and B.~Barbara, {\it ``Quantum tunneling of magnetization in molecular 
complexes with large spins and effects of environment''}, {\it Magneto-Science - From Molecules to
Materials}, Edited by M.~Drillon and J.~Miller, (Wiley VCH Verlag Gmbh, 2001).

\bibitem{ref02} L.~Landau, \JL{Phys.~Z.~Sowjetunion,2,1932,46}; C.~Zener, 
\JL{Proc.~R.~Soc.~London,A137,1932,696}; E.C.G.~Stuckelberg, \JL{Helv.~Phys.~Acta,5,1932,369}.

\bibitem{ref03} S.~Miyashita, \JPSJ{64,1995,3207}; L.~Gunther, \JL{Euro.~Phys.~Lett.,39,1997,1}; 
V.V.~Dobrovitski and A.K.~Zvezdin, \JL{Euro.~Phys.~Lette.,38,1997,377}.

\bibitem{ref04} A.J.~Leggett, S.~Chakravarty, A.T.~Dorsey, M.P.A.~Fisher, A.~Garg, W.~Zwerger,
\JL{Rev.~Mod.~Phys.,59,1987,1}; M.~Grifoni, P.~Hanggi, \JL{Phys.~Rep.,304,1998,56}; Y.~Kayanuma, 
H.~Nakayama, \PR{B57(20),1998,13099}; E.~Shimshoni and A.~Stern, \PR{B47,1993,9523}.

\bibitem{ref05} A.~Abragam, B.~Bleaney, {\it Electronic Paramagnetic Resonance of Ions}, (Clarendon 
Press, Oxford, 1970) (Chapter 10).

\bibitem{ref06} A.L.~Barra et al., \JL{J.~Am.~Chem.~Soc.,114,1992,8509}. 

\bibitem{ref07} D.~Gatteschi et al., \JL{Molecular Engineering,3,1993,157}.

\bibitem{ref08} I.~Chiorescu, PhD Thesis, University Joseph Fourier, Grenoble (2000).

\bibitem{ref09} I.~Chiorescu, W.~Wernsdorfer, A.~Muller, H.~Bogge, and B.~Barbara, 
\PRL{84,2000,3454}.

\bibitem{ref10} B. ~Barbara, L.~Thomas, F. ~Lionti, A. ~Sulpice, A.~Caneschi, 
\JL{J.~Magn.~Magn.~Mat,177,1998,1324}.

\bibitem{ref11} M.I.~Katsnelson, V.V.~Dobrovitski,and B.N. ~Harmon, \PR {B159,1999,6919}.

\bibitem{ref12} J.H.~van Vleck, \PR{59,1941,724}. 

\bibitem{ref13} I.~Chiorescu, W.~Wernsdorfer, A.~Muller, H.~Bogge, and B.~Barbara, 
\JL{J.~Magn.~Magn.~Mat,221,2000,103}.

\bibitem{ref13b} G.~Chaboussant~et~al., cond-mat 0204365.

\bibitem{ref14} R.~Giraud, W.~ Wernsdorfer, A.M.~Tkachuk, D.~Mailly, and B.~Barbara, 
\PRL{87,2001,057203-1}.

\bibitem{ref15} H.A.~Kramers, \JL{Proc.~Amsterdam~Acad.,33,1930,959}.

\bibitem{ref16} F.D.M.~Haldane, \PRL{93,1983,454}; ibid \andvol{50,1983,1153}.

\bibitem{ref17} Y.~Ajiro et al. to be published.

\bibitem{ref18} S.~Miyashita and N.~Nagaosa, \PTP{106,2001,533}.

\bibitem{ref19} R.~Harris, J.~Wiley and sons, \JL{Magnetic Resonance in Spectroscopy,6,1986,100}.

\bibitem{ref20} K.~Saito  and S.~Miyashita, \JPSJ{70,2001,3385}.

\end{thebibliography}
\end{document}